\documentstyle[12pt]{article}
\catcode`\@=11
\begin{document}

\def \A {{\cal A}}
\def\Jo#1#2#3#4{{#1} {\bf #2}, #3 (#4)}
\def \re#1{(\ref{#1})}
\def\st{\scriptstyle}
\def\sst{\scriptscriptstyle}
\def\mco{\multicolumn}
\def\epp{\epsilon^{\prime}}
\def\vep{\varepsilon}
\def\ra{\rightarrow}
\def\vp{{\bf p}}
\def\al{\alpha}
\def\ab{\bar{\alpha}}
\def \bi{\bibitem}
\def \ep{\epsilon}
\def\D{\Delta}
\def\sms{$\s$-models }
\def \om {\omega}

\def\be{\begin{equation}}
\def\ee{\end{equation}}
\def \lab {\label}
\def \k {\kappa} 
\def \F {{\cal F}}
\def \g {\gamma}
\def \del {\partial}
\def \bd {\bar \partial }
\def \na {\nabla}
\def \const {{\rm const}}
\def \ha{{\textstyle{1\over 2}}}
\def \na {\nabla }
\def \D {\Delta}
\def \a {\alpha}
\def \b {\beta}
\def \chi {\chi}\def\r {\rho}
\def \s {\sigma}
\def \p {\phi}
\def \m {\mu}
\def \n {\nu}
\def \vp {\varphi }
\def \l {\lambda}
\def \t {\theta}
\def \td {\tilde }
\def \d {\delta}
\def \ci {\cite}
\def \la {\label}
\def \sm {$\s$-model }
\def \foot {\footnote }
\def \P {\Phi}
\def \o {\omega}
\def \inv {^{-1}}
\def \ov {\over }
\def \four{{\textstyle{1\over 4}}}
\def \fourth{{{1\over 4}}}
\def \foot{\footnote}
\def\be{\begin{equation}}
\def\ee{\end{equation}}
\def\bea{\begin{eqnarray}}
\def\eea{\end{eqnarray}}
\def\np {{\em  Nucl. Phys. }}
\def \pl {{\em  Phys. Lett. }}
\def \mpl {{\em Mod. Phys. Lett. }}
\def \prl {{ \em  Phys. Rev. Lett. }}
\def \pr  {{\em  Phys. Rev. }}
\def \ap  {{\em Ann. Phys. }}
\def \cmp {{\em Commun.Math.Phys. }}
\def \ijmp {{\em Int. J. Mod. Phys. }}
\def \jmp {{\em J. Math. Phys.}}
\def \cqg {{\em  Class. Quant. Grav. }}

%%%%%%%%%%%%%%%%%%%%%%%%%%%%%%%%
%%%%%%%%%%%%%%%%%%%%%%%%%%%%%%%%%%
%\begin{document}
%\title{EXTREME DYONIC BLACK HOLES IN STRING THEORY}
%\author{ A.A. TSEYTLIN }
%\address{Theoretical Physics Group, Blackett Laboratory\\
%Imperial College, London SW7 2BZ, U.K.}
%%%%%%%%%%%%%%%%%%%%%%%%%%%%%%%%%%%%%%%%%%%%%%%%%%%%%%%%%%%%%%
%\maketitle
%\abstracts
\begin{titlepage}
\begin{flushright}Imperial/TP/95-96/22\\hep-th/9601177\\
January (revised February) 1996\\
 Mod. Phys. Lett. A11 (1996) 689

\end{flushright}
\vskip 3cm
\begin{center}
{\Large\bf Extreme  dyonic black holes   }\\
\vskip 0.2cm
{\Large\bf  in string theory}
\vskip 1.5cm
{\bf A.A. Tseytlin\footnote{e-mail: tseytlin@ic.ac.uk. \  On leave from Lebedev Institute, Moscow.}}\\
\vskip 0.2cm
{\it Theoretical Physics Group, Blackett Laboratory }\\
{\it Imperial College,   London SW7 2BZ, U.K.}
\end{center}

\begin{abstract}
{Supersymmetric extreme dyonic black holes of toroidally 
compactified  heterotic or type II string theory 
can be viewed as lower-dimensional
images of  solitonic strings 
wound around a compact dimension. 
We consider conformal sigma models which describe 
string configurations  corresponding to various 
extreme dyonic black holes in four and five dimensions. 
These conformal models have  regular 
short-distance region equivalent to  a  
WZW theory with level proportional to magnetic charges.
Arguments are presented suggesting a universal relation between 
 the black hole  entropy (area) and  the 
statistical entropy of BPS-saturated oscillation states 
of solitonic string.}
\end{abstract} 
\end{titlepage}

\newpage 
%%%%%%%%%%%%%%%%%%%%%%%%%%%%%%%%%%
\section{Introduction}
%%%%%%%%%%%%%%%%%%%%%%%%%%%%%%%%%%
String theory generalises 
the Einstein theory in many ways.
In addition to the Einstein term,  the low-energy effective action of string theory 
contains  also a  combination 
of extra massless  `matter' 
 fields with special couplings;  there are   classical higher derivative
$\a'$-corrections; exact  classical 
string solutions are described by conformal 
field theories which often correspond to higher dimensional backgrounds and  their properties  should be determined by studying  
 propagation of test strings instead of test particles,  etc.
As a result, string theory introduces several 
new aspects in the black-hole physics. At the same time, 
it is  also  likely  to  provide
answers to some  old questions.

In general, properties of leading-order  
string solutions are modified  by both  classical 
($\a'$) and quantum ($e^\P$)  string corrections.
There exists a special class of 
 extreme  (supersymmetric, BPS saturated) 
black hole solutions for which 
these corrections are expected (see e.g.  \ci{kall})
 to be  more  under control
and which,  hopefully,  may be used as  `solvable models' 
to study some   aspects of black hole physics.

The  extreme  field configurations 
contain vector fields of Kaluza-Klein origin
and thus  are actually 4-dimensional `images' (or dimensional 
reductions) of higher dimensional backgrounds. 
It turns out that electrically charged  extreme $D=4$ black holes
are, in fact,  approximate descriptions 
 of higher (e.g. $D=5$) dimensional 
fundamental string backgrounds \ci{HRT}. 
The latter are field configurations produced by classical 
fundamental  strings wound around compact spatial 
directions, i.e. they are supported by 
$\d$-function sources \ci{duh}. This provides an explanation  \ci{CMP,dab}  of 
why  the electrically charged BPS saturated 
 extreme black holes are in one-to-one correspondence 
with BPS states in elementary string spectrum in flat space
  \ci{DRU,seen}.

The condition of 
BPS saturation is still 
 satisfied if, in addition to    some winding and momentum numbers  
(related to electric charges of $D=4$ black hole), 
the  fundamental string  source 
is  also  oscillating 
in  compact directions (e.g.,  in  the left-moving sector
in the heterotic string case)  \ci{duh,CMP,dab}.
Higher-dimensional backgrounds produced by 
fundamental strings with different oscillation patterns  
reduce to   a   family of  extreme  black holes  
with the same  electric  charges 
but different non-vanishing {massive} 
 `Kaluza-Klein' fields  (`hair') 
 which 
are invisible   at scales larger than the compactification scale.
This  
suggests   a natural way of  understanding  the 
thermodynamic black-hole entropy  
in terms of degeneracy of string configurations  \ci{suss, sen} 
which  give rise to 
 black holes  with  the same values of asymptotic
 charges  \ci{CMP,dab}.
To  reproduce the black hole entropy
as a statistical entropy  $\ln d(N)$ it  is important that 
the oscillating object  is  a string-like, i.e. has 
 exponentially growing  number  of states 
at a given oscillator level $N$
(it is not enough to consider just a Kaluza-Klein field theory).\foot{Though singular at the classical level 
the fundamental string solutions should become regularised 
(get $\sqrt{\a'}$-`thickness') once quantum oscillations 
 of the source (i.e. $\a'$-corrections) are taken into account \ci{ME}. This 
expectation 
is  essentially equivalent in the present context to
 the suggestion  \ci{sen}
that the thermodynamic black-hole  entropy, which 
vanishes when evaluated
at the singular  horizon  ($r=0$)
 of the leading-order effective field theory solution,
should instead be computed at the `stretched'  
 horizon  \ci{suss} at $r \sim  \sqrt {\a'}$.
The resulting expression  is then  proportional to  
 the statistical  entropy  corresponding to free string states  \ci{sen,peet}.}

The  magnetically charged  extreme black holes 
are dimensional reductions  of  certain $D\geq 5$    solitonic configurations \ci{HMON,kalor}. They are regular, i.e.   have a  finite size of the  `core'.\foot{The  magnetic charge
plays the role of a short-distance regulator, 
 providing an effective shift   $r \to r +P$ 
(analogous to  $r \to r+ \sqrt {\a'}$  that  should  happen 
in the exact fundamental string solution).}
While for the extreme electric black
 hole or fundamental string  solution the 
 effective string coupling (exponential of the dilaton, $e^{2\P} = e^{2\P_\infty}(1 +{Q\ov r})^{-1}$) 
goes to   zero  at small scales, 
 it blows up near the origin for the magnetic solution 
($ e^{2\P} = e^{2\P_\infty}(1 + {P\ov r})$). 
As a result,  string loop
corrections  may not  be ignored.

Remarkably, this problem can be avoided
if  one  considers certain
 extreme dyonic black holes. The condition of preservation of 
supersymmetry  
implies  \ci{kall,CY} that the electric ($Q$) and magnetic ($P$)
 charges 
should correspond to different  vector fields.
In addition, 
since we are interested in string-theory solutions which have
exact conformal \sm interpretation,  
  these vector fields (at least part of them)
should originate from 
 different internal dimensions.\foot{Supersymmetric
dyonic solutions  
with  $Q$ and $P$ corresponding to the two vector fields 
 ($G_{5\m}$ and $B_{5\m}$) associated with the same Kaluza-Klein 
dimension  \ci{KOO,IWP} 
can be obtained from 
purely-electric  or purely-magnetic  solutions
by $SL(2,R)$ duality rotations  
(which  change \sm metric and in general may  not preserve manifest conformal invariance of 
$\s$-model since $S$-duality is not a symmetry at string tree level).
 Dyonic backgrounds 
with both $Q$ and $P$ associated with the same vector field
(see, e.g., \ci{KHO}) completely 
break supersymmetry \ci{CY}  and  do not correspond 
to a manifestly conformal 2d theory, i.e. are  non-trivially 
deformed by $\a'$ corrections.
I am grateful to T. Ort\'{\i}n for clarifying discussions of related 
issues.}
These black holes  are  reductions of   higher ($D \geq 6$) dimensional   string solitons \ci{US,USS} with all of the 
background fields  being  regular.
As in  the purely  electric case
there is actually  a large class of $D=4$ dyonic BPS-saturated 
black hole backgrounds which differ
only by their short-distance structure
and correspond to BPS-saturated excitations 
of the solitonic string \ci{LW,USS}.

The dilaton field is 
generically given by 
 a  ratio of  the magnetic and  electric   charge factors,  e.g.,  $e^{2\P} = e^{2\P_\infty}  (r + {P})/(r + {Q})$.  
The approximate constancy of the dilaton  both at large and small distances
ensures that these four-dimensional  dyonic black holes
are more `realistic' than purely electric or purely magnetic ones. Indeed, these 
solutions have   global space-time structure of  extreme 
Reissner-Nordstr\"om  black holes.
Also, in contrast to the purely  electric or magnetic  extreme 
black holes which have 
zero area of  the horizon,  here the area is proportional 
to the product of electric and magnetic charges \ci{CY}, opening a possibility to 
 compare the 
statistical entropy to Bekenstein-Hawking one  \ci{LW}.
By analogy with the corresponding counting for purely electric  black holes (oscillating fundamental string  counting)
  \ci{sen,CMP,dab}
the entropy  should now have  an interpretation in terms
of counting of  different BPS-saturated  
 solitonic string states with  given  values of asymptotic 
 charges  \ci{LW,USS}.

Below we shall consider an   example  of  such background 
 described by   magnetically charged 
(via coupling to  sixth internal dimension)
solitonic string wound around fifth compact dimension
(with winding and momentum numbers giving rise to electric charges
of the $D=4$ black hole)  \ci{US}.
The important point is that here  one is dealing with explicitly known 
   well-defined conformal $\s$-model.  
  For large values of  charges
the level of the WZW-type conformal field theory
which describes the  short-distance  (`throat') region
is large and thus the counting of  excited states should 
proceed  in  more or less the  same way  as  in flat space case 
up to a rescaling  of the string tension by a  product of magnetic charges \ci{USS}, as 
suggested in \ci{LW}.
As a result, one   is able to 
reproduce  the Bekenstein-Hawking expression for the black hole 
entropy 
using   semiclassical
string-theory considerations \ci{LW,USS}.
The universality of this approach is confirmed by a similar 
analysis of the case of (both non-rotating and rotating)
 $D=5$ extreme dyonic black holes 
(the special 
case of  $D=5$ dyonic black holes with 
 all charges being equal  
was  considered using  different methods in \ci{SV,MV}).

We shall start with  listing  the  conformal \sms 
 describing string configurations which 
represent, upon dimensional reduction, 
 to  various $D=4$ and $D=5$ extreme black hole
 backgrounds  (Section 2).
We shall then  concentrate on  the simplest  models (corresponding to 4-charge dyonic black hole in four dimensions \ci{CY,US} 
 and 3-charge
dyonic black hole in five dimensions)
and  explain, both for $D=4$ and $D=5$ cases, 
 how one may relate  their thermodynamic 
entropy  to the statistical entropy  of
excited  BPS-saturated solitonic string
 states   (Section 3). 
In Section 4 we shall consider a conformal \sm  describing 
a  rotating  $D=5$ supersymmetric extreme dyonic black hole
which generalises the solution in \ci{MV}. 
We shall find that  conditions of conformal invariance 
and regularity at short distances demand  the 
presence of two equal components of rotation in two orthogonal 
planes. The consistency of 2d conformal model implies 
a quantization condition for the angular momentum and a bound on its maximal value.

Most of the results  about  $D=4$ dyonic black holes
have already appeared in  \ci{US,USS}  while the  construction 
of $D=5$ dyonic black holes  and  
statistical understanding  of their 
entropy using  the  conformal \sm approach  are  new.

%%%%%%%%%%%%%%%%%%%%%%%%%%%%%%%%%%%%%%%%%%%%%%%%%%%%%%%%
\section{Conformal models for  fundamental and solitonic strings,
  and   extreme black holes}
%%%%%%%%%%%%%%%%%%%%%%%%%%%%%%%%%%%%%%%%%%%%%%%%%%%%%%%%%
Classical string solutions are described  by conformal 2d $\s$-models.
In the case when the background fields (\sm couplings) do not depend on some  compact isometric  coordinates  one may re-interpret 
the \sm action  as representing a lower-dimensional 
background with extra Kaluza-Klein 
fields (for a review see, e.g., \ci{TS}).

Below we shall discuss the conformal models 
for  certain  fundamental and solitonic 
string backgrounds, which, upon dimensional reduction, 
may be identified  as $N \geq 1$ supersymmetric 
$D=4$ or $D=5$  extreme black holes.
We shall  give  only the bosonic parts 
of the corresponding $(2,1)$  (in type II theory case)
or $(2,0)$ (in heterotic string theory case)
world-sheet supersymmetric 
Lagrangians.

%%%%%%%%%%%%%%%%%%%%%%%%%%%%%%%%%%%
\subsection{Fundamental strings and electric black holes}
%%%%%%%%%%%%%%%%%%%%%%%%%%%%%%%%%%%%%%%%%%%%
The fundamental string solutions are described by the 
$D=10$  conformal chiral null models  \ci{TH} with flat transverse part.
An example is provided by ($u=t+y, \ v=t-y$)
\be
   L =  F(x)  \del u \left[\bd v +
   K(x) \bd u \right]
 +   \del x^i \bd x^i    +    {\cal R}
\P (x)\  .  
 \la{lag}
\ee
This model  is  conformal  to all orders if 
\be
 \del^2 F\inv=0, \ \ \ \del^2 K\inv=0, 
\ \ \  \P =  \P_\infty + \ha \ln F \ . 
\la{harr}
\ee
If the functions depend on  $D-2$
 transverse dimensions,   then 
\be
F\inv = 1 + {Q_2 \ov r^{D-4}}, \ \ \ \ 
K= K_0 + {Q_1 \ov r^{D-4}}, \ \ \ \ \  r^2=x^i x^i  , 
\la{FKK}
\ee
describes  (for $K_0=0$)  a 
background produced by a fundamental string  \ci{duh,dab}
wound around the compact dimension 
$y\equiv y+2\pi R$  with the winding number $w$ and the momentum number 
$m$ along $y$  being 
 proportional to $Q_2$ and $Q_1$, 
\be
Q_1=  {16 \pi G^{(D-1)}_N \ov  (D-4) \om_{D-3}}\cdot {m \ov R}\ ,
\ \ \ \ \ 
Q_2=  {16\pi  G^{(D-1)}_N \ov  (D-4) \om_{D-3}}\cdot {wR \ov \a'}\  . 
\la{ree}
\ee
Here $\om_{D-3} = {2\pi^{{D-2 \ov 2}}/ \Gamma ({D-2\ov 2})}$
is the area of unit sphere in $D-3$ dimensions
and $G^{(D-1)}_N = G^{(D)}_N/2\pi R$  is the Newton's constant in $D-1$ dimensions (which is fixed under $T$-duality,
$m \leftrightarrow w, \  R \leftrightarrow \a'/R, \ 
Q_1 \leftrightarrow Q_2$).\foot{We shall use the following notation.
  The free  string action
is   $I= (\pi \a')\inv \int d^2\s \del x \bd x $ $ 
=(4\pi \a')\inv$ $ \int d^2\s \del_a x \del^a x$.
The  string  effective action   is 
$S_D=(16\pi G^{(D)}_N)\inv $ $\int d^{D} x 
$ $  \sqrt {-G} e^{-2 \P'} (R + ... )$=
$(16\pi G^{(D)}_N)\inv$ $  \int d^{D} x 
 \sqrt {-g}  (R_g + ... )$.  Here $G_{\m\n}$  and $g_{\m\n}$ 
are  the  string-frame  and Einstein-frame metrics (both  equal  to $\eta_{\m\n}$ at $r\to \infty$), $ \P'$ is the non-constant part of the  dilaton (i.e. $ \P'_\infty =0$).
%=\P- \P_\infty - \four {\rm det} G_{int}$).  
In general, the Newton's
constant  is $G^{(D)}_N \sim   e^{2 \P_\infty}\a'^4/V_{10-D}$ where 
$ V_{10-D}$ stands for  the volume of internal 
$(10-D)$-dimensional space.}
In the case of $D=5$ fundamental string  $x^i=(x^s,y^n)$ where $x^s$ ($s=1,2,3)$
are three non-compact  spatial coordinates and  $y^n $  ($n=1,...,5$) are  toroidally
compactified coordinates  (with $ y\equiv y^6$).
 The  dimensional reduction   along 
$y$ gives   extreme electrically charged  $D=4$ black hole. 
Equivalent $D=4$ background is found by 
choosing $K_0=1$,  $ u = y', v = 2t $  and 
dimensionally reducing 
along the `boosted' coordinate $y'$. Starting with the  $D=5$  model \re{lag},\re{ree}  with
\be
F\inv = 1 + {Q_2 \ov r}, \ \ \ \ 
K= 1 + {Q_1 \ov r}, \ \ \ \ \  r^2=x^s x^s  ,  
\la{FK}
\ee
\be
Q_1=  {4G_N}{ m \ov  R } \equiv 4G_N \bar Q_1,
\ \ \ \ \ 
Q_2=  {4 G_N }{ R w \ov  \a' }\equiv 4G_N \bar Q_2, 
\ \ \  \  G_N \equiv G^{(4)}_N , 
\la{qeq} 
\ee
one finds  the  following 
$D=4$  string-frame metric \ci{TH}
\be
 ds^2 = - \l^2 (x) dt^2 + dx^sdx^s \ , \ \ \  \l^2=  FK\inv  \ .  
\la{stt}
\ee
The Einstein-frame metric  and the (non-constant part of) $D=4$ dilaton  are
\be
ds^2_E =- \l (r)  dt^2 + \l\inv (r)  (dr^2 + r^2
d\Omega^2_2) \ ,   \la{eee}
\ee
\be
 \l  (r) = e^{2 \P'_{(4)}}= {r\ov \sqrt
 {(r + {Q_1})(r + {Q_2 })} }\ . 
\la{ele}
\ee
In addition, there are two vector fields and a scalar modulus
(radius of $y$-direction).
The $T$-duality in the $y$-direction interchanges 
the `winding' and `momentum' charges $Q_2$ and $Q_1$
(i.e. interchanges the vector fields associated with 
$B_{\m y}$ and $G_{\m y}$).

The special cases of $Q_2=0$ and  $Q_1=0$ 
describe the Kaluza-Klein ($a=\sqrt3$)
 extreme electric black hole  and  its $T$-dual, 
while the case of $Q_1=Q_2$  corresponds to the  dilatonic
($a=1$) extreme black hole  \ci{GIBMA}.

Other supersymmetric BPS-saturated  black hole solutions
with the same values of electric charges $Q_1,Q_2$
but different short-distance structure are found
by starting with a   more general 
conformal model 
(which can be viewed as an  integrated marginal deformation 
of the model \re{lag})
describing fields produced by 
 BPS-saturated oscillating string states of the free string spectrum 
with fixed values of the winding and momentum numbers 
 in the $y$-direction.
This model  is itself a special case of the  chiral null model \ci{TH}
 \be
   L =  F(x)  \del u \left[\bd v +
   K(u,x) \bd u  +   2{\cal A}_i(u,x)  \bd  x^i \right]
 +    \del x^i \bd x^i    +    {\cal R}
\P (x)\  ,  
 \la{lagd}
\ee
where,  for conformal invariance, 
${\cal A}_i$ should satisfy $\del_i \F^{ij}=0, \ \F_{ij} = \del_i {\cal A}_j - \del_j {\cal A}_i$. 
 For example, one may choose 
$K= f_n (u) y^n + {Q_1/ r}, \ {\cal A}_5={q / r} $
where $f_n$ will then be  related to the profile of `left'  
oscillations  of the string source  
in `transverse' directions $y^n$ ($\del_u^2 X^n =-2 f^n$)
and $q$ will be the (`left') electric charge of the string.
The level matching condition for the string source
will    then imply $Q_1Q_2 - q^2 \sim < (\del_u X^n)^2>$ \ci{dab}.
Viewed from 4 dimensions (i.e. after averaging in $u$), this background will
describe a family of   extreme electrically charged 
black holes with the same  electric charges $(Q_1,Q_2;q,-q)$
(corresponding to 6-th and 5-th internal dimensions)
but different short-distance structure  (massive `hair')
depending on a choice of the oscillation  profile  function \ci{CMP,dab}.

%%%%%%%%%%%%%%%%%%%%%%%%%%%%%%%%%%%%%%%%%%%%%%%%%%%%%%%%%%
\subsection{Superconformal $\s$-models  and magnetic black holes}
%%%%%%%%%%%%%%%%%%%%%%%%%%%%%%%%%%%%%%%%%%%%%%%%%%%%%%%%%%
The magnetic counterparts of the extreme $D=4$ electric black holes
can be found by  $D=4$  $S$-duality transformation, 
or by considering the above fundamental string model 
as a 6-dimensional background and applying   the  $S$-duality
transformation in 6 dimensions ($ G \to  e^{-2\P} G$,  \ 
$ dB \to  e^{-2\P} \ast dB, $
\ $\P \to -\P$).\foot{This is a formal symmetry of the part of $D=6$ heterotic or type II effective action which does not include extra vector fields.}
At the string-theory 
 level the corresponding solitonic configurations
are described by $N=4$ superconformal models with the bosonic part of the Lagrangian given by \ci{US} 
$$
L=  -\del t \bd t +  \del y^n \bd y^n  + 
 f(x)k(x)  \big[ \del y + a_s (x) \del x^s\big] \big[ 
\bd y  + a_s (x) \bd x^s\big] 
    $$    
\be
 +\  f(x)  k^{-1} (x) \del x^s \bd x^s  + 
  b_s (x) (\del y \bd x^s - \bd y \del x^s)
  +   {\cal R}   \p(x) \  .    
 \la{lara} 
\ee
As above,   $x^s$ are three non-compact spatial dimensions,
while  $y$ and $y^n$  are compact coordinates. To have 
$N=4$  ($(4,0)$ in the heterotic  or $(4,1)$ in the type II case) world-sheet supersymmetry and conformal invariance  we shall assume 
that  all the fields  depend only on  $x^s$  and $f,k,a_s,b_s,\p$ 
are subject to
\be
   \del_s\del^s f =0 \ ,  \ \ \ \ \ \ \ \ \ \ \ \    \del_s\del^s k\inv =0 \ ,  
\lab{hepp}
\ee
\be
 \del_{p} b_q - \del_q b_{p}
 =- \epsilon_{pqs} \del^s   f   , \ \ \ 
\del_{p} a_q - \del_q a_{p}
 = -\epsilon_{pqs} \del^s   k\inv , \ \  \ \p= \ha \ln f . 
\lab{hypp}
\ee
Under $T$-duality in $y$-direction $f \to k\inv ,  \   k \to f\inv , 
\  a_s \to  b_s ,   \ b_s \to  
a_s  ,$  (provided  $a_{[p} b_{q]}=0$).
The simplest 1-center choice for the harmonic functions
\be
f = 1+{P_2\over r},\ \ \  
k\inv = 1+{P_1\over r} ,  \ \ \  a_s dx^s= P_1 (1-\cos \t) d\vp , \ \  b_sdx^s = P_2 (1-\cos \t)d\vp,
\lab{fk}
\ee
leads, upon dimensional reduction along $y$ and $y^n$ directions, 
to the $D=4$  extreme  magnetic black hole backgrounds parametrised by the two charges $P_1$ and $P_2$. The 
special cases of this model include:  the 
Kaluza-Klein monopole or 
 $a=\sqrt 3,$  $D=4$ magnetic black hole ($P_2=0$), its $T$-dual -- the $H$-monopole  background  \ci{HMON} ($P_1=0$), 
and  the $T$-self-dual model 
corresponding to the $a=1$ dilatonic magnetic black hole  \ci{kalor}($P_1=P_2$).
A generalization of \re{lara} to the case of several
magnetic charges (obtained by adding extra free coordinates and applying $T$-duality transformation)  was discussed in \ci{kalorr}.

The  $D=4$ metric corresponding to \re{lara} has the form  (\ref{stt}),(\ref{eee}) with
\be
 \l  (r) = \sqrt {kf\inv} = e^{-2 \P'_{(4)}}= {r\ov \sqrt
 {(r + {P_1})(r + {P_2 })} }\ . 
\la{magt}
\ee
As in the case of the `5-brane' solitonic model  \ci{chs},  the 
short-distance  $r\to 0$  region  of the model \re{lara} 
is  a regular `throat' and is described by (a factor of) $SU(2)$ WZW theory
with level $\k= 4P_1P_2/\a'$.
%%%%%%%%%%%%%%%%%%%%%%%%%%%%%%%%%%%%%%%%%%%%%%%%%%%%%%%%%%%%
\subsection{Solitonic strings  and  dyonic  black holes}
%%%%%%%%%%%%%%%%%%%%%%%%%%%%%%%%%%%%%%%%%%%%%%%%%%%%%%%%%%%%
It  may look  rather surprising that  
the `electric' and `magnetic' models \re{lag} and \re{lara} 
can be directly superposed while preserving the exact conformal invariance  and  $N=2$ world-sheet  
 ($N=1, D=4$ space-time  \ci{CY}) supersymmetry. 
The resulting  conformal model  (with non-trivial 6-dimensional part) is  defined  by \ci{US} 
$$
   L =  F(x)  \del u \left[\bd v +
   K(x) \bd u \right] +  f(x)k(x)  \big[ \del y_1 + a_s (x) \del x^s\big] \big[ 
\bd y_1  + a_s (x) \bd x^s\big] 
$$  
\be
 +\  f(x)  k^{-1} (x) \del x^s \bd x^s  + 
  b_s (x) (\del y_1 \bd x^s - \bd y_1 \del x^s)
+ \del y^k \bd y^k   +   {\cal R}   \P(x) ,   
\la{larr} 
\ee
$$
 \P= \P_\infty + \ha \ln (F f) \ , 
$$
 where $u=y_2, v=2t,$  $k=3,4,5,6$, all functions depend
only on $x^s\ (s=1,2,3)$ and $F\inv,K,f,k\inv$ are harmonic 
as in \re{harr},\re{hepp},\re{hypp}.
With the functions chosen as in \re{FK},\re{fk} this model  describes
a magnetically charged  solitonic string wound around
a compact `fifth' dimension.\foot{Some special cases and related models were discussed also  in \ci{behh,behhh}.}

The corresponding
$D=4$ background 
obtained by dimensional reduction along six  compact $y$-coordinates
is the  supersymmetric BPS-saturated 
extreme dyonic 
black hole  (first found as a leading order  $D=4$ 
solution of $T^6$-compactified heterotic string in  \ci{CY}).\foot{It is possible, of course, to consider more general (e.g. multi-center)
choices for the harmonic functions $F\inv,K,f,k\inv$.
In particular, taking all 4 functions to be $O(3)$-symmetric but
having centers at different points 
one finds  the solution  discussed in \ci{RRR}.}
In addition to the 2 electric and 2 magnetic vector fields
and two scalar moduli (radii of $y_1$ and $y_2$ circles)
the dyonic background includes   the  Einstein-frame metric \re{eee} with   
\be
 \l  (r) = \sqrt {FK\inv kf\inv}= {r^2\ov \sqrt
 {(r + {Q_1})(r + {Q_2 })(r + {P_1})(r + {P_2 })} }\ , 
\la{mag}
\ee
and the  dilaton $\P$  or  the effective  $D=4$ dilaton $\P'_{(4)}$
\be
e^{2\P} = e^{2 \P_\infty}\sqrt
 {{r + {P_2}\ov r + {Q_2 }} }\ , \ \ \  \ 
 e^{2 \P'_{(4)}} = \sqrt{ {(r + {P_1})(r + {P_2 }) \ov (r + {Q_1})(r + {Q_2 }) }} \ .
\la{dii}
\ee
Assuming all charges are positive, $r=0$ is a regular horizon
(for $P_i=0$ the horizon  coincides with singularity). 
There  is also 
 a
time-like  singularity  at  
 $r_{sing}=-{\rm min}\{P_1,P_2,Q_1,Q_2\}$,  
i.e.
 the global structure of $D=4$  space-time is that  of
extreme Reissner-Nordstr\"om black holes \ci{CY}.\foot{If all charges are equal
the dilaton is constant and the solution coincides with 
the extreme dyonic Reissner-Nordstr\"om black hole
of  the standard Einstein-Maxwell ($a=0$) theory.
The case of $Q_1=Q_2=P_1, \ P_2=0$ corresponds to  a solution in the case of  
dilatonic coupling  constant $a=1/\sqrt3$ \ci{CY,DLR}.}
It should be noted that these conclusions apply if one considers this 
background as a usual 4-dimensional one and  applies the 
  standard rules of geodesic
extension 
 (what corresponds to analytic continuation 
 of  $r$ to negative values).
At the same time, if such  
backgrounds are 
considered  as string-theory solutions, they  are effectively  
higher-dimensional when probed at small scales and thus 
their short-distance structure  should  actually  be analyzed  from  a higher dimensional  point of view 
(more precisely, from the point of view of underlying conformal field theory).   The extension of  $r$ to negative values 
seems  not to make sense in that case.

The ADM mass of black hole and the area of the event horizon 
 (${\bf A}$ $\equiv$ $ 4\pi(\l\inv  r^{2})_{r=0}$)
are
 found to be
\be
M= {1\ov 4 G_N}(Q_1+Q_2+P_1+P_2)\ , \ \ \  \ \ 
\ \ 
{\bf A}= 4\pi \sqrt{ Q_1 Q_2 P_1 P_2} \ . 
\la{maar}
\ee
The expression for the mass can be   written  also as 
\be
M = \bar Q_1 + \bar Q_2 + g^{-2} (\bar P_1 + \bar P_2)
 , \  \ 
\la{maaa}
\ee
where  we have chosen $G_N\equiv G^{(4)}_N={1\ov 8}\a' e^{2 \P_\infty}
\equiv  {1\ov 8} \a' g^2$ 
 and  $
Q_i \equiv  {4G_N } \bar Q_i$ $ 
  =  \ha \a'g^2 \bar Q_i, $  \   
$ P_i \equiv  \ha \a' \bar P_i$ (cf.\re{qeq}). 
One needs all four charges to be 
non-vanishing to get a non-zero area 
of the horizon and thus a non-zero  value of the  analogue 
of the Bekenstein-Hawking entropy\foot{Extremal black holes 
have zero temperature and  simplest examples of them were shown
to  have zero thermodynamic entropy \ci{ENT}.
This  zero entropy conclusion does not, however, apply in the 
present 4-charge case as one can show by first defining the 
entropy in the non-extremal case 
(the corresponding solution was given in  \ci{CY})   
and then taking the extremal 
limit.  It is important that in contrast to the simple 
Reissner-Nordstr\"om case  here the limit involves several 
different parameters.
Thus  the Bekenstein-Hawking formula  is  still  expected to apply
 to  the case of  these extreme dyonic black holes (cf.\ci{gho}).
I am grateful to  M. Cveti\v c  for 
 explanations  on  this point.}
\be 
{\bf S}_{BH}\equiv  {{\bf A} \ov 4G_N}  \ . 
\la{ent}
\ee
As in the  purely electric case, there is a whole family of 
 black hole solutions  with the same values of electric and magnetic charges but different short-distance structure.
They are 4-dimensional `images' of  certain  (supersymmetric BPS-saturated) excited  
states of the solitonic string  \re{larr} which are  described by  a marginal `deformation'
of   \re{larr} which includes extra `chiral' couplings
(see \re{lagd} and the next subsection). 
The knowledge of  underlying 2d conformal  field theory 
makes possible to count the number  $d(N)$ of such states
for a given set of (large) charges $(Q_1,Q_2,P_1,P_2)$
and to  identify  \ci{USS}  the Bekenstein-Hawking entropy \re{ent}
with the statistical entropy $\ln d(N)$, 
supporting  the proposal of  \ci{LW}.
This will be   discussed  in  Section 3.

%%%%%%%%%%%%%%%%%%%%%%%%%%%%%%%%%%%%%%%%%%%%%%%%%%%%%%%%%%%%
\subsection{Generalisations and related models}
%%%%%%%%%%%%%%%%%%%%%%%%%%%%%%%%%%%%%%%%%%%%%%%%%%%%%%%%%%%%
The models \re{lara},\re{larr}  
are special cases of  a  chiral null model with 
curved transverse part\foot{When  both $K$  and ${\cal A}_i$ 
depend on $u$ one can redefine $v$ to absorb $K$ into ${\cal A}_i$ 
\ci{TH}.} 
\be
   L =  F(x)  \del u \left[\bd v +
   K(u,x) \bd u  +   2{\cal A}_i(u,x)  \bd  x^i \right]
 +  
\ha {\cal R} \ln F(x) 
+ L_{\bot}\ , 
\la{laggg}
\ee
\be
L_{\bot}= (G_{ij} + B_{ij})(x) \del x^i \bd x^j    +    {\cal R}
\p (x)\  . 
 \la{lggg}
\ee
There exists a renormalisation  scheme  
in which (\ref{laggg}) is conformal to all 
orders in $\a'$ provided
  the `transverse'  \sm \re{lggg}  is conformal
   and  the functions 
$F\inv,K,{\cal A}_i, \P$ satisfy
certain  conformal invariance conditions.
The simplest tractable case is when the transverse
theory has at least $(4,0)$ 
extended  world-sheet supersymmetry so that the conformal invariance
conditions  essentially preserve their 1-loop form  \ci{HP}, i.e. 
are the `Laplace'  equations 
in the  `transverse' background 
(for simplicity  here we  assume that $K$ and $\A_i$ are  $u$-independent)  \ci{TH,USS}
\be
 \del_i   (e^{-2\p} \sqrt G G^{ij} \del_j F^{-1}) =0 \  ,  \ \ 
\ \ \ \del_i   (e^{-2\p} \sqrt G G^{ij} \del_j K) =0 \  ,
\la{LL}
\ee
\be
 \hat  \nabla_i (e^{-2\p} {\cal F}^{ij} ) =  {1 \ov \sqrt G}  \del_i   (e^{-2\p} \sqrt G {\cal F}^{ij})  -
\ha e^{-2\p}  H^{kij} {\cal F}_{ki}
=0  \  ,    
 \la{cond}
\ee
 where $\hat \Gamma^i_{jk}
=  \Gamma^i_{jk} + {1\over 2} H^i_{\ jk}  ,  \ \ 
 {\cal F}_{ij} \equiv  \partial_i {\cal A}_j -
\partial_j {\cal A}_i .$

For example,  the 8-dimensional transverse  space may be 
chosen as a direct product  of some 4-space $M^4$ and a 4-torus.
If the  functions defining $M^4$-model  have $SO(3)$ symmetry 
  we are led back to the case of \re{larr} with 1-center harmonic functions.
Choosing ($s=1,2,3$) 
\be
{\cal A}_i(x)  \bd  x^i 
= A(x)[\bd y_1 + a_s (x)  \bd x^s]\ , \ \ \ \ \ 
A= {q\ov r} \cdot { r +  \ha (P_1+P_2) \ov r + P_1}\ , 
\la{gene}
\ee
we find a generalisation \ci{USS}  of the `dyonic' model \re{larr}
which  describes  a spherically symmetric $D=4$ background with two extra 
electric charges $(q,-q)$ and the metric function and  the area of  event horizon being (the mass remains the same as in \re{maar})
\be 
\l={{r^2}\over{\sqrt{(r+Q_1)(r+Q_2)(r+P_1)(r+P_2) 
 -  q^2 [r + \ha(P_1 +P_2)]^2} }} \   , 
\la{meto}
 \ee
 \be
{\bf A} 
= 4\pi \sqrt{ Q_1Q_2P_1P_2 - \four  q^2(P_1
+ P_2)^2} \ 
.  \la{mes}
\ee
Further generalisations  of the dyonic model  \re{larr} 
(responsible for 
the `hair' of 4-dimensional dyonic backgrounds) 
are obtained, e.g.,  by  introducing  $\A_i$ 
which describes rotation and/or Taub-NUT charge (see 
also \ci{TH} and  Section 4)
or by switching on the $u$-dependence in $K$ 
(or, equivalently, in $\A_i$) as in  the fundamental string  case \re{lagd} discussed above.

Another possibility is to consider 6-dimensional models 
with $M^4$-part  having (locally)  $SO(4)$-invariant structure.
In this case the transverse theory 
 is the same as in the 5-brane  model \ci{chs}
\be
L_{\bot}= f(x) \del x^m \bd x^m +   B_{mn }(x)  \del x^m \bd x^n  + 
{\cal R}
\phi (x)\  ,  \ \ \ \p= \ha \ln f ,
\ \ \del^2  f =0,  
\lab{hppp}
\ee
$$
    G_{mn} = f(x) \delta_{mn} \  ,  \ \ \ \ 
 H_{mnk} = - 2\sqrt G G^{pl} \epsilon_{mnkp} \del_l \p 
= - \epsilon_{mnkl} \del_l  f ,
$$
where   $m,n, ...=1,2,3,4$.\foot{The  $D=10$ model \re{laggg},\re{hppp}
may thus  be considered as an anisotropic generalisation 
of the 5-brane  model of \ci{DL,chs}:
different isometric 5-brane coordinates are multiplied by different functions of the 4 orthogonal coordinates,  cf.\ci{MK}.
In a sense,   it describes   a fundamental string 
`lying' on a 5-brane:   
 the  5-brane  is wrapped around  4-torus times $S^1$ ($u$-circle)
with  the fundamental string   wound  around the latter. 
The non-singular solitonic nature of the 5-brane is responsible for regularity  of the  `combined'  background.}
One  finds that the Laplace-type  equations for $F\inv, K$ preserve their  flat-space form \re{harr}
and thus (we assume that 
${\cal A}_i=0$; cf.\re{FK},\re{fk})
\be
f= 1 +{P\ov r^2} , \ \ \ 
F\inv = 1 +{Q_2\ov r^2} , \ \ \
K= 1 +{Q_1\ov r^2} , \la{ooo}
\ee
$$
e^{2\P} = Ff= {{ r^2 + P}\over{r^2 + Q_2}}  , \ \ \ \ 
\  \  r^2\equiv x^m x^m. 
$$
The special cases of this model are:
(1) six-dimensional fundamental string ($P=0$)
with momentum  and winding numbers $\sim Q_{1,2}$ \ci{duh}; 
(2) its $D=6$  $S$-dual -- solitonic 
string  solution \ci{dukh} ($Q_1=Q_2=0,P\not=0$) 
which  is   a   $D=6$  reduction 
 of $D=10$  5-brane solution \ci{chs,DL};
(3) `dyonic' $D=6$ string  of \ci{duf} ($Q_1=0, Q_2\not=0, P\not=0$).
A generalisation  of this background containing   one extra  (angular momentum)
parameter  will be constructed in Section 4.

Dimensionally reducing the corresponding   $D=6$ model to 5 dimensions along $u$ (or, from the point of view of the corresponding $D=10$ solution, 
 wrapping the 5-brane around $S^1\times T^4$) one finds  the 3-charge $(Q_1,Q_2,P)$  extreme dyonic $D=5$ black hole background.\foot{In five  dimensions  an antisymmetric tensor is dual to a vector so that $H_{mnp}$ 
is a   magnetic dual  of an  electric field with charge $P$.
The duality   maps a  $D=5$ solution with 2 electric and one `magnetic'  charge  of one theory into a solution with 3 electric charges of a dual theory.}
The 5-dimensional string-frame  metric  and
the
Einstein-frame metric  are related by $g_{\m\n} = e^{-{4\ov 3 }  \P'_{(5)}} G_{\m\n}$, $ e^{4 \P'_{(5)}}=  f^2 {FK^{-1}}, \ e^{2\P}=e^{2 \P_\infty} Ff,$ 
so that 
 \be
ds^2_E =- \l^{2} (r)  dt^2 +  \l\inv  (r)  (dr^2 + r^2 d\Omega^2_3) \ , 
\la{eeee}
\ee
$$
\l =  (F\inv K f )^{-1/3}  = {r^{2} \ov [(r^2+Q_1)(r^2+Q_2)(r^2+ P)]^{1/3}}\ . 
$$ 
The mass  of this $D=5$ black hole, the 3-area of the regular 
 $r=0$  horizon 
and the Bekenstein-Hawking entropy are (cf.\re{maar},\re{ent})\foot{In general, for a black hole in  $D$ dimensions with 
$g_{tt} = -1 + {\mu/r^{D-3}} + ...$ 
the ADM mass is (see, e.g., \ci{MPE}) \  $M= \m (D-2)\om_{D-2}/16 \pi G^{(D)}_N$
\ ($\om_2=4\pi, \ \om_3=2\pi^2$, etc.). 
 The Bekenstein-Hawking entropy is expressed in terms of the volume $A$ of $(D-2)$-dimensional horizon surface 
by ${\bf S}_{BH} = {\bf A}/ 4G^{(D)}_N$.}
\be
 M=  {\pi \ov 4G_N} (Q_1 +Q_2 + P)  ,  \ \ \ \ \  \ \
G_N \equiv G^{(5)}_N \ ,  
\la{maa}
\ee
\be
{\bf A}  =  2 \pi^2 \sqrt {Q_1Q_2P}  \ , \ \ \  \ \ \ 
{\bf S}_{BH} = {{\bf A} \ov 4G_N} \  . 
\la{amaa}
\ee
If  $Q_1=Q_2=P$
we find $\l\inv  = f =1 + {P / r^2}, \ \P= \P_\infty$. Introducing $\r^2=r^2 + \r^2_0, \ \r_0\equiv  \sqrt P$
we  then finish with the $D=5$ extreme 
 Reissner-Nordstr\"om  metric 
\be
ds^2_E =- \l^2   dt^2 +  \l^{-2}  d\r^2 + \r^2 d\Omega^2_3 \ , \ \ 
\ \ \ \
 \l = 1 - {\r_0^2/ \r^2}  \ . 
\la{eeew}
\ee
This  special case ($Q_1=Q_2=P$)  
was discussed in  \ci{SV} in connection with reproducing the Bekenstein-Hawking entropy as a statistical entropy  using  the  D-brane  approach to count
the number of  corresponding microscopic BPS states.
As in the case of the 
$D=4$ dyonic black hole model,  this counting can be done  
also in a direct way 
(for generic  large  values of  $(Q_1,Q_2,P)$) 
using  the  fact that the small-scale  (`throat') region
 of the  corresponding $D=6$ conformal 
model is described by a WZW model with level 
$\kappa = P/\a'$ (see Section 3.3).

There are also many other exact  $D=10$ superstring solutions described by \re{laggg},\re{lggg}. For example, we may take the 8-dimensional 
transverse model  to be a   product of the two 
`5-brane' models \re{hppp}
 \be
   L =  F(x,y)  \del u \left[\bd v +
   K(x,y) \bd u  \right]
 +  f_1(x) \del x^m \bd x^m  +   f_2(y) \del y^m \bd y^m
\la{ygg}
\ee
$$
+ \   B_{1mn}(x) \del x^m \bd x^n  +   B_{2mn}(y) \del y^m \bd y^n
 + \ha {\cal R} \ln[ F(x,y) f_1(x)f_2(x)]
   \ , 
$$
 where $f_1$ and $f_2$ are  harmonic.
The conformal invariance conditions 
$\nabla^i(e^{-2\p}\del_i F\inv)=0,$  $\nabla^i(e^{-2\p}\del_iK)=0$
reduce to\foot{I am grateful to J. Maldacena for pointing out a mistake 
in the equation below in the original version of this paper.}
$$
 [ f_2(y)  \del^2_x +  f_1(x) \del^2_y] F\inv (x,y) =0 \ , 
\ \ \ \ 
[ f_2(y)  \del^2_x +  f_1(x) \del^2_y] K (x,y) =0 \ , 
$$
 and thus are solved, e.g.,  by
\be
F(x,y) = F_1(x)F_2(y), \ \ \  \  \  K(x,y) = K_1(x) K_2(y),
\la{gwg}
\ee
where $F_{1,2}$ and $K_{1,2}$ are independent harmonic functions.
For  the simplest 1-center  choice of the functions   this  $D=10$ `dyonic' background is parametrised by 6 charges. The special case  ($K=1$) of this solution 
was found  in \ci{duf}.

%%%%%%%%%%%%%%%%%%%%%%%%%%%%%%%%%%%%%%%%%%%%%%%%
\section{Dyonic black hole entropy from string theory}
%Degenerate black hole states and statistical entropy}
%%%%%%%%%%%%%%%%%%%%%%%%%%%%%%%%%%%%%%%%%%%%%%%%

The $D=4$ dyonic black holes  (with four \re{mag} or five \re{meto}  parameters) discussed above  belong to  the set 
of $N=1$ supersymmetric BPS saturated extreme dyonic solutions of the leading-order effective equations of  $T^6$-compactified  heterotic string 
which are parametrised by  28 electric and 28 magnetic charges \ci{CY,USS,CYY}.\foot{These more general $D=4$ solutions can be constructed \ci{USS,CYY}
by applying special $T$- and $S$-duality transformations
to the 5-parameter solution but not all of them  directly 
correspond to exact 
string solutions, i.e. to  manifestly conformal  $D=10$ 
$\s$-models.}  
 For fixed values  of charges $(Q_i,P_i)$ 
one expects to find a subfamily of $N=1$ supersymmetric BPS-saturated  black hole backgrounds
which all look  the same at large distances but  differ 
in their short-distance structure (i.e. at scales of order of compactification scale where their higher dimensional solitonic string origin becomes apparent). This  classical `fine structure' or `degeneracy' 
should  be responsible for the black hole entropy \ci{CMP,dab,LW}.

To try to check  the proposal \ci{LW} that one can indeed reproduce the Bekenstein-Hawking entropy  as a statistical entropy of degenerate black hole configurations it is sufficient to consider the simplest  
non-trivial choice  of the charges ($Q_1,Q_2,P_1,P_2)$.
The aim is to  explain 
the expression for the corresponding  entropy \re{ent}
in terms of degeneracy of BPS states
originating from possible 
small-scale oscillations of underlying 
six-dimensional string soliton  \re{larr},\re{FK},\re{fk}.
The relevant `oscillating' $D=6$   backgrounds  are described by the  general model \re{laggg} and can be thought of as
 special marginal deformations 
of \re{larr}.
Since  \re{larr} defines  a non-trivial CFT, the counting is not as simple  
 as in  the fundamental string (`pure electric') case, where,  
because of the  matching onto string sources,  one expects 
that  
 BPS-saturated oscillating  fundamental string configurations 
 are in one-to-one 
correspondence with  analogous  BPS states in the free string spectrum  
\ci{CMP,dab}.   Still,  it  can be done at least  approximately 
(for large charges) 
in a rather straightforward way.

The key observation is that since 
 the degeneracy is  present only at small scales (all corresponding $D=4$ black hole backgrounds look the same 
at large $r$)  to  find the number of different states
one may  first replace  \re{larr} by its short-distance ($r\to 0$)
limit and then do the counting. 
In this `throat'  limit  \re{larr} reduces to a WZW-type  model 
with level $\k=4P_1P_2/\a'$. Then  for large values 
of the level  the counting of  BPS states should  proceed
 essentially as in the  free string case, with   only  two important 
 differences. One 
is that now the  string tension of the {\it `transverse'}
part of the action 
is  proportional to $\k$.
Another is that in contrast  to what happens in the free string case,
here the number of BPS oscillation  states  we should  count 
is the {\it same} in the  heterotic and type II theories. Indeed, 
 the relevant 
marginal perturbations  of the model \re{larr}
 are only `left', not `right' (the functions in \re{laggg}
can  depend only on $u$, not on $v$ to preserve the conformal invariance
when both $F$ and $K$ are non-trivial).
This is also related to the fact that the
 background in \re{larr}
is `chiral' and has the same amount of space-time supersymmetry ($N=1,D=4$)
in both theories (only `left' perturbations will be  supersymmetric). 
As a result, the entropy  should  be the same in 
heterotic and type II cases, in agreement with the fact that 
the corresponding black hole solutions 
(and thus the Bekenstein-Hawking  entropy)
are the same.

This approach is legitimate if {\it all}  four charges ($Q_1,Q_2,P_1,P_2)$ 
are large compared to the compactification scales (radii  $R_n$ of compact  coordinates $y_1,y_2$ which we shall assume to be of order of $\sqrt{\a'}$). 
Indeed, the scale of the soliton is determined by $\sqrt{P_1P_2}$
and thus is large (i.e. the curvature at the throat is small)
provided  the magnetic charges are large.
 At the same time,  if the  large-distance value of string coupling
is  taken to be small ($e^{\P_\infty} <1$), 
 to ensure that its short distance  (`throat') value
 ($e^{\P_{0}}$) is also small (see \re{dii}),   
one is to assume that  the electric charges  are of the same order as 
the magnetic ones.
Similar remarks apply to the case of the $D=5$ dyonic  black hole model 
discussed in Section 2.4.
%%%%%%%%%%%%%%%%%%%%%%%%%%%%%%%%%%%%%%%%%%%%%%%%%%%%%%%%%%%%
\subsection{Throat region model and magnetic
 `renormalisation' of string tension}
%%%%%%%%%%%%%%%%%%%%%%%%%%%%%%%%%%%%%%%%%%%%%%%%%%%%%%%%%%%%
The model \re{larr},\re{FK},\re{fk} has a regular $r\to 0$ limit
described by \ci{US,USS}
\be
 I= {1\ov \pi \a'} \int d^2 \s  L_{r \to 0}  =  {\k \ov 4\pi }\int d^2 \s   \ (\del z \bd z + \del \td  u  \bd \td  u 
 +    e^{-z }   \del \td  u \bd \td v   
\la{ttr}
\ee
$$
 +   \    \del \td  y_1 \bd \td  y_1   +  \del \vp \bd \vp + 
\del \theta \bd \theta  - 2 \cos \theta  \del  \td  y_1 \bd \vp )  $$
$$= \  {1 \ov \pi\a' }\int d^2 \s  \left( e^{-z }   \del   u \bd  v  + Q_1  Q_2\inv \del   u  \bd   u  \right) 
$$
$$
  + \ 
{\k \ov 4\pi }\int d^2 \s  \left(\del z \bd z
 +  \del \td  y_1 \bd \td  y_1   +  \del \vp \bd \vp + 
\del \theta \bd \theta  - 2 \cos \theta  \del  \td  y_1 \bd \vp   \right) . $$
Here 
%($u\equiv  y_2, \ v\equiv 2t$) 
 \be   \kappa= {4\ov \a'}  P_1P_2 \ , \ \ \ \ \  \  z \equiv \ln {Q_2 \ov  r}   \to \infty \  ,
\la{not}
\ee
$$   \td  u =(Q_1\inv  {Q} _2 P_1P_2 )^{-1/2}  u, \ \ \ \td v= (Q_1 Q_2\inv P_1P_2)^{-1/2} v , \ \ \ 
\td  y_1 = { {P}_1\inv }  y_1 + \vp  .
$$
As already mentioned above, an 
important property of this model is that 
here (in contrast to, e.g.,   the 5-brane model
 \ci{chs}) the dilaton is {\it constant} 
in the wormhole region.\foot{Assuming that all 4 charges  are of the same order
we may ignore the  difference between  the values of the  dilaton (string coupling)
at  $r=\infty$  and at $r=0$ (see \re{dii}).}

The  throat region model (\ref{ttr}) 
is   equivalent to a direct product 
of   the   $SL(2,R)$ and $SU(2)$ WZW  theories
(divided  by discrete subgroups) with  both levels 
   equal to $\k$.\foot{Similar model (corresponding to the special case of $P_1=P_2,\ Q_1=Q_2$) was previously discussed in \ci{SL}.}
Since the level of  $SU(2)$  must be integer, we get 
the quantisation condition for the product of magnetic charges,  
$P_1P_2 = \four \a' \k  $.\foot{Moreover, 
each of $P_1$ and $P_2$ should also be quantized.
For example, the requirement of regularity of the 
6-metric $\sim [dy_1 + P_1(1-\cos \t )d\vp]^2 + ...$
    (the absence of Taub-NUT 
singularity) implies that  
one should be able to identify $y_1$  
with  period $4\pi P_1$,  which is possible if  $2P_1/R_1 = k$. 
By   $T$-duality in $y_1$-direction
  the same constraint  should apply   also to 
$P_2$, i.e.  $2P_2= l  \td R_1= l\a'/R_1$. }

For large values of charges, i.e. large $\k$, 
one may expect that 
the counting of  states in this model may  proceed 
essentially as if this was a 
 flat space case.
 The only  (but crucial) difference is 
that the   {\it transverse} ($z,\td y_1, \vp, \theta$) part
of the action  has now 
 {\it `renormalised'} string tension
 %(we set the radii $R_n$ of compact 
%dimensions to be  equal to $\sqrt {\a'}$) 
\be  
{1\ov \a' }\   \to \  {1\ov \a'_\bot } 
= {P_1P_2\ov \a'^2} \ . 
\la{tens}
\ee
Though  this rescaling  seems to be the correct final result, 
to think that counting  can be  done  exactly   as in flat space is an oversimplification:
the presence of non-vanishing electric charges is actually as important 
as the presence of the magnetic ones.
This is crucial  to take into accout in order to demonstrate that 
{\it all}  the transverse coordinates 
get renormalised tension.\foot{I am grateful to F. Larsen for raising this issue.} 
While  the coefficients in front of 
`longitudinal' terms $\del u \bd v$ and $\del u \bd u$
can be rescaled, 
this does not apply to the coefficient of $\del z \bd z$ term
(which determines the level of $SL(2,R)$).
Since  $(u,v,z)$ are  coupled, the  information about the 
scale of $\del z \bd z$
 `propagates' into the rest of the transverse part of the action  (omitted in \re{ttr}) through 
the dependence of perturbations on  $u$.
The relevant (supersymmetric, left-moving) 
perturbations  in all other free transverse compact 
coordinates are  represented  by the $\A_i$-term in \re{laggg}:
\be L'= L +  V,  \   \ \ \ \  \ \  V=  2F(x) \A_n (u,x) \del u \bd y^n \ .
\la{aap}
\ee
They are marginal if $\A_n = q_n(u)/r $ where $q_n$ are related to 
  profiles of oscillations in $y_n$ directions. These perturbations vanish  at large $r$ while near $r=0$  we get 
$V\approx  2q_n(u)Q_2\inv  \del u \bd y^n$.
The dependence on magnetic charges enters indirectly -- 
 due to the fact that $u$ is coupled to $z = \ln (Q_2/r)$. For $r\to 0$ we 
get (cf.\re{ttr})
\be
L' =  e^{-z} \del u \bd v  + Q_1Q_2\inv \del u \bd u + 
P_1P_2  \del z \bd z  + 2q_n(u)Q_2\inv  \del u \bd y^n
 +   \del y_n \bd y_n  + ... \ ,
\la{ooop}
\ee 
where dots stand for the 3-sphere and constant dilaton terms.
To see the effect of non-vanishing $q_n$ let us integrate out $y_n$.
Then 
\be 
L'=  e^{-z} \del u \bd v 
    +   c(u) \del u \bd u 
    + P_1P_2  \del z \bd z  + ... \ , 
\ \ \ \  c(u)\equiv Q_2^{-2} [Q_1Q_2  - q_n(u) q_n(u)] \ . 
\la{ijn}
\ee
Note that all transverse  `charges' $q_n(u)$ enter  on an equal footing.
Since $(u,v,z)$-theory is $SL(2,R)$ WZW model 
with level $\k = 4{\a'}^{-1} P_1P_2 \gg 1 $
the  coefficient in front of $<q_nq_n >$ term  in the stress tensor  or Virasoro operators  will thus be  rescaled  by  a factor of $1/\k$. At the same time, $<q_nq_n >$
should have  an  interpretation of the oscillator level.
 Solving the classical equations for $v,z$  using $1/\k$  expansion
(assuming  that $z$ is finite, i.e.  that $\k$ goes  
to infinity  before $r$ goes to zero)
  and computing the Virasoro 
operators  one finds that 
for a winding (in $u$) string  state  
the level matching condition becomes $mw \sim {1 \ov \k} <q_nq_n >$
or $N_L \sim  <q_nq_n > \sim \k mw$.

More explicitly, using the flat-space counting 
picture  one finds that  if $w$ and $m$  are the winding and momentum 
 numbers of a free  heterotic string  compactified on a circle of radius $R$, 
then the mass and oscillator level  of  BPS-saturated  left-moving  oscillation 
 modes ($\hat N_R=0, \ \hat N_L=N_L - 1$) are given by the standard expressions  (cf.\re{maar},\re{maaa})
\be 
M = {m\ov R} + {wR\ov  {\a'}}
= {1\ov 4G_N} (Q_1 +Q_2)\ , \ \ \   \ \ 
\hat N_L = mw = {\a' \ov 16 G_N^2}Q_1 Q_2\ , 
\la{hete}
\ee
where we have used  \re{qeq} to express $m,w$ in terms of  the 
`space-time' charges $Q_1,Q_2$. The relations
\re{hete} are true in the case when  both  $S^1$  and non-compact 
terms in the string 
 action
have the same overall coefficient (tension).
 The  magnetic renormalisation  \re{tens} of the tension 
of the transverse part of the action  implies that $N_L$ is to be 
rescaled by the ratio of the `longitudinal' ($1/2\pi\a'$)
and `transverse' ($1/2\pi\a'_\bot$) tensions (the oscillator level is proportional to the inverse of string tension)
\be
N_L \to N_L = {\a'\ov \a'_\bot} mw =\four \k mw =  {P_1P_2 Q_1 Q_2 \ov 16 G_N^2} \  . 
\la{ttt}
\ee
Since the charges are large we ignore  the quantum `$-1$' shift.

%%%%%%%%%%%%%%%%%%%%%%%%%%%%%%%%%%%%%%%%%%%%%%%%%%%%%%%%%%%%
\subsection{Statistical entropy}
%%%%%%%%%%%%%%%%%%%%%%%%%%%%%%%%%%%%%%%%%%%%%%%%%%%%%%%%%%%%
The number of   BPS states in the free string spectrum
with a given  left-moving oscillator number $N_L\gg 1$   is  
(see, e.g., \ci{suss})
$d(N_L)_{N_L\gg 1} \approx  c_0  N_L^\nu  \exp(4\pi \sqrt {N_L})$.
Then the statistical entropy  of  an ensemble of states 
with the same  charges but different  left-moving oscillation modes
is given by 
\be
{\bf S}_{stat}=  \ln d(N_L)_{N_L \gg 1}  \    \approx \  {  4 \pi }  \sqrt { N_L  }  \ 
.
\la{mesti}
\ee
Using the expression  \re{ttt}  for $N_L$ 
we then find  that, for large charges, 
  ${\bf S}_{stat}$ coincides with  the Bekenstein-Hawking entropy \re{ent},  
\be
{\bf S}_{stat} \approx \  {  4 \pi }  \sqrt { N_L  }  
 =   {  2 \pi }  \sqrt {\k m w  } = 
  {\pi \sqrt{ P_1 P_2 Q_1 Q_2} \ov  G_N}  = {{\bf A}\ov 4G_N} = {\bf S}_{BH}
\ . 
\la{een}
\ee
It should be emphasized that there is no ambiguity 
in the overall coefficients in these expressions for the  entropy.
First, there is no  ambiguity 
in the coefficient in front of the thermodynamic entropy 
(in contrast to the  stretched horizon approach  
to   the purely electric  case  where  the numerical coefficient 
in front of ${\bf S}_{BH}$ depends on a choice of a position of the stretched horizon \ci{sen,peet}).
Also, the coefficient in front of $\sqrt {N_L}$ in $d(N_L)$ 
is related to  the fact that underlying degrees of freedom 
correspond to a (`half' of) 1-dimensional extended object, i.e. are described by a 
2d  field theory (see also \ci{LW}).  
Note also that the dependence on the  Newton's constant 
effectively drops out.\foot{The  
statistical entropy is given by  a classical string-theory expression 
and does not depend on  the string coupling. 
The constant $G_N$ needs to be introduced to express   (cf.\re{qeq})  the string  `microscopic' 
quantum numbers in terms of `space-time' charges
(which appear in the effective field theory description). This is 
necessary in order 
to be able to compare with the Bekenstein-Hawking entropy. Since 
${\bf S}_{BH}$ itself 
originates from the on-shell value of the (euclidean)
space-time effective action it also  contains a factor of $1/G_N$.}

%%%%%%%%%%%%%%%%%%%%%%%%%%%%%%%%%%%%%%%%%%%%%%%%%%%%%%%%%%%%
\subsection{$D=5$  extreme dyonic black holes }
%%%%%%%%%%%%%%%%%%%%%%%%%%%%%%%%%%%%%%%%%%%%%%%%%%%%%%%%%%%%
The universality of the relation between the statistical 
and Bekenstein-Hawking entropies  
is confirmed  by analogous  consideration of the 
case of $D=5$  extreme dyonic black holes 
described by \re{laggg},\re{hppp}--\re{amaa}.
Here the  throat limit $r\to 0$ is described
by a similar   $SL(2,R) \times SU(2)$ 
WZW model (cf.\re{ttr}--\re{tens})\foot{The `transverse' ($z, \psi, \vp, \theta$) part
 of the  throat region   is exactly the  same as in the  5-brane model 
 \ci{chs}  (where $Q_1=Q_2=0$) except for the fact that 
here the dilaton $\P$ 
is { constant}
in the $r\to 0$  region, i.e. the string coupling is not blowing up.}
\be
 I_{r \to 0}  
=  {1 \ov \pi\a' }\int d^2 \s  \left( e^{-z }   \del   u \bd  v  + Q_1  Q_2\inv \del   u  \bd   u  \right) 
\la{nnn}
\ee
$$
  + \ 
{\k \ov 4\pi }\int d^2 \s  \left(\del z \bd z
 +  \del \psi \bd   \psi  +  \del \vp \bd \vp + 
\del \theta \bd \theta  - 2 \cos \theta  \del  \psi \bd \vp    \right)  ,  $$
\be   \kappa= {1\ov \a'}  P \ , \ \ \  \  z \equiv \ln {Q_2 \ov  r^2}   \to \infty \  , \ \ \ \ \ 
  {1\ov \a'_\bot } 
= {P\ov 4\a'^2} \ . 
\la{note}
\ee
Its  integer level $\k = P/\a'$    again 
rescales the tension of the `transverse' part of the action
by 
  $\a'/\a'_\bot= P/4\a'$.
The important factor of 4 difference compared to the expressions in \re{not},\re{tens}  is related to the  increased dimensionality of the  spatial sphere (here the harmonic functions
depend on $r^2$; note also that the standard metric on $S^3$ with unit radius is $ds^2 = \four[d\t^2 + \sin^2 \theta d\vp^2 + (d\psi - \cos \theta d \vp)^2]$, where $0 \leq \t \leq \pi, \   0 \leq \vp
 \leq 2\pi,\ 
0 \leq \psi \leq 4\pi$). 

The analogues of the relations \re{hete},\re{ttt} are found by 
 using  the expressions for the charges 
(we apply \re{qeq}  for $D=6$ fundamental string; as in \re{maa} here    $G_N \equiv  G^{(5)}_N$)
 \be
Q_1= {4\ov \pi} G_N  \cdot { m\ov R}\ , \ \ \ \ Q_2= {4\ov \pi}  G_N  \cdot {wR\ov \a'}\ , \ \ \ \   P= \k \a'\  , 
\la{chaa}
\ee
\be 
M = {m\ov R} + {wR\ov  {\a'}}
= {\pi \ov 4 G_N} (Q_1 +Q_2)\ , \ \ \   \ \ 
\hat N_L = mw = {\a' \pi^2 \ov 16 G_N^2}Q_1 Q_2\ , 
\la{hetep}
\ee
\be
N_L \to N_L = {\a'\ov \a'_\bot} mw = \four k mw =  { \pi^2  P\ov 64 G_N^2}  Q_1 Q_2\ . 
\la{tttp}
\ee
The expression for the statistical entropy is then (cf.\re{amaa},\re{een})
\be
{\bf S}_{stat} \approx \  {  4 \pi }  \sqrt { N_L  }  
 =   {  2 \pi }  \sqrt {\k m w  } = 
 {\pi^2 \sqrt{ P Q_1 Q_2 } \ov  2 G_N}  = {{\bf A}\ov 4G_N} = {\bf S}_{BH} .  
\la{eeen}
\ee
Notice again a remarkable  consistency  of numerical factors.

In the special case of $Q_1=Q_2=P$ the same conclusion 
was reached in \ci{SV} by  first 
transforming  the  heterotic string  solution into
a  type II string  one using  the $D=6$  $S$-duality and  then 
 counting  the number of corresponding  BPS states as
 $D$-brane bound states.
As the  approach (described in \ci{USS}  and  above)   
 based directly on the underlying conformal field theory, 
 this 
  `$D$-brane counting'  approach  also uses  the 
assumption of large  charges  and also  
 reduces  to 
evaluation of  a number of    
states  in  some 2d conformal (hyperk\"ahler)  $\s$-model.
A simpler  $D$-brane counting  derivation of the entropy
of $D=5$ extreme  dyonic black holes 
 which applies to  the general case  of arbitrary large charges 
$Q_1,Q_2,P$ was  recently given in \ci{CM}.
%%%%%%%%%%%%%%%%%%%%%%%%%%%%%%%%%%%%%%%%%%%%%%%%%%%%%%%%%%%%%
%%%%%%%%%%%%%%%%%%%%%%%%%%%%%%%%%%%%%%%%%%%%%%%%%%%%%%%%%%
 \section{Rotating $D=5$ extreme dyonic black holes}
%%%%%%%%%%%%%%%%%%%%%%%%%%%%%%%%%%%%%%%%%%%%%%%%%%%%%%%%%%%5
The conformal 
chiral null model \re{laggg} with non-vanishing $\A_i$-term
describes also  rotating  purely electric extreme 
supersymmetric  black holes  
in various dimensions. Indeed, \re{laggg} with 
flat transverse part  represents  \ci{TH}  IWP backgrounds \ci{IWP}
which include, in particular, extreme Taub-NUT and rotating 
solutions (for axi-symmetric choices of harmonic functions).
The corresponding extreme solutions (with rotation in only one plane) 
have naked ring singularity
in $D=4,5$ but  get  regular  horizon 
and saturate Bogomol'nyi bound in $D \geq 6$ \ci{HS}.
More complicated  rotating solutions in $D=4$ (which   have the
same large $r$  asymptotics   as  the extreme supersymmetric 
low-energy solutions 
obtained by applying $T$-duality transformations to Kerr solution  \ci{seen}  but do not have 
naked singularities)   can be constructed by  taking a
periodic array  of  $D>5$  black holes \ci{HS}
or by starting with the chiral null model \re{laggg}
with $u$-dependent coupling $K$ or $\A_i$ 
(which describes a background produced by oscillating or 
rotating  $D=5$  fundamental string) and averaging  
over the compact direction \ci{dab} (in this latter 
case there is a natural 
Regge bound on the maximal value of the angular momentum 
but the short-distance form of the  solution is not explicit).

One may wonder how the properties of $D=4,5$  rotating solutions 
change once one adds magnetic charges, i.e. 
considers extreme supersymmetric
dyonic black holes described by \re{laggg}
with curved transverse part.\foot{The importance of this problem was emphasized to me by M. Cveti\v c. I would like to thank her 
for helpful discussions of several aspects of rotating solutions.}
It turns out that the rotating  version  of the $D=4$ 
background  corresponding  to  the model 
\re{larr} still has a ring singularity as in  the pure electric case.\foot{The presence of  the singularity may, in  principle, 
be harmless at the string-theory  level provided the corresponding 
5-dimensional conformal \sm  is regular.}
At the same time, there exists a  natural 
rotating generalisation of the 3-charge   $D=5$ 
dyonic model \re{ooo},\re{eeee}
which describes a non-singular rotating 
dyonic black hole solution of 
$D=10$ heterotic or type II theory compactified on $S^1 \times T^4$.
In the special case when all 3 charges are equal 
this model is a conformal \sm behind 
 the solution found in \ci{MV}
(which itself is a rotating version of the solution of \ci{SV}).

  As we shall see below,  the requirement of conformal invariance
of underlying \sm 
(which  reduces to a condition of self-duality of the strength of 4-potential $\A_i$ in \re{laggg})
implies that (as in \ci{MPE,MV}) 
 one needs {\it two}  equal components of rotation
 in the two orthogonal planes. 
A remarkable feature of this model is that like  in the absence 
of rotation  (Section 3.3) 
it has a {\it regular} throat region described by  a similar
$SL(2,R) \times SU(2)$  WZW model.  The  bound on the
 maximal value of the angular momentum and its quantisation  then follows
directly from the 
conformal  field theory  considerations.\foot{This solution
  has a non-singular horizon only if all 3 charges are non-vanishing. 
A generalisation of extreme purely electric  $D=5$ 
solution in \ci{HS} to the case of two  components of rotation 
remains singular.}
The expression for the entropy of the resulting 
rotating black hole can again be understood in terms of counting of 
possible BPS deformations of the  conformal model
(a $D$-brane counting derivation of the entropy in the special 
equal-charge case was given in \ci{MV}). 

%%%%%%%%%%%%%%%%%%%%%%%%%%%%
\subsection{$D=6$ conformal $\s$-model and $D=5$ black hole}
%%%%%%%%%%%%%%%%%%%%%%%%%%%%%%%%
Let us consider  a 6-dimensional  model \re{laggg} 
with  the 
 `transverse' $M^4$-part  $(G_{mn},$ $ B_{mn},$ $  \p)$ having 
torsion related to dilaton in the following specific way
($ m,n,...=1,2,3,4$)
\be
 H^{mnk} = - {2 \ov  \sqrt G}  \epsilon^{mnkl} \del_l \p   \ . 
\la{torr}
\ee
This  relation is satisfied  both  when the $M^4$-part
is represented by  $SO(4)$-invariant   
 `5-brane'   $\s$-model \re{hppp} and
  in the case of $SO(3)$-invariant model in  \re{larr}
(there one needs to assume that $y_1=x_4, \ x_s=(x_1,x_2,x_3)$ 
and that $a_s \del_s \p=0$ which is satisfied for the  
 1-center background \re{fk}).
 Then 
the  conformal invariance  condition for 4-vector $\A_m$ 
 \re{cond}   takes  remarkably simple  form
\be    
\del_m   (e^{-2\p}  \sqrt G  {\cal F}^{mn}_+ ) =0   \ ,
\la{sell}
\ee
$$
 \F^{mn}_+  \equiv  {\cal F}^{mn} + {\cal F^*}^{mn} \ , \ \ \ \ \ \ 
{\cal F^*}^{mn} = {1\ov 2 \sqrt G  }  \ep^{mnkl} {\cal F}_{kl}  \ . 
$$
The simplest solutions  existing  for 
arbitrary  $\p$    
are thus  given by abelian  4-dimensional instantons 
\be
(\sqrt G G^{mn} G^{kl} +  \ha \ep^{mknl} ) {\cal F}_{nl} = 0  \ . 
\la{ins}
\ee
Since this  (anti)selfduality equation does
 not depend on a scale of $G_{mn}$, 
in  the  $SO(4)$-invariant case of \re{hppp} (corresponding  to the $D=5$ solution
discussed below) 
it does not depend on the  `magnetic'  function $f$,
 i.e. takes the {\it flat space} form.
 Choosing  the 4 spatial coordinates  as 
$ x^1+ix^2 = r \sin \theta e^{ i \vp },$  
\  
$ x^3 + ix^4=  r \cos \theta  e^{ i \psi },$
\be
 dx^m dx_m  =
 dr^2 + r^2 (d\theta^2 +
 \sin^2\theta d\vp^2  + \cos^2\t d\psi^2)\   , 
\la{ccr}
\ee
the equation \re{ins}  can be solved by  a
 simple ansatz   with 
rotational symmetry in  the two orthogonal planes
\be
 \A_\vp = \A_\vp(r,\t)  , \ \ \  \  \A_\psi  = \A_\psi(r,\t)  , \ \ \ 
\ \  \A_r=\A_\theta=0  , 
\la{ansa}
\ee
\be
   \F^{r\vp}_+ = { r \cot \t \ \del_r \A_\vp - \del_\t \A_\psi \ov r^3 \sin \theta \cos \theta} 
 \ , \ \ \ \ 
 \F^{\t\vp}_+ = {r\inv  \cot \t \ \del_\t \A_\vp +  \del_r \A_\psi \ov r^3 \sin \theta \cos \theta}\  ,  
\la{FFF}
\ee
$$
\F^{\theta\psi}_+  = - r\inv \tan \t \ \F^{r\vp}_+ \ , \ \ 
\ \ \ \F^{r\psi}_+  =  r \tan \t \  \F^{\t \vp}_+  \ . 
$$
Imposing  $\F^{mn}_+ =0$ one finds\foot{In addition,  there is a growing solution 
$ \A_\vp = \g  { r^2} \sin^2 \t  , \ 
\A_\psi = - \g { r^2} \cos^2 \t , $
which describes a rotating magnetic universe
(a similar solution  with  one rotational plane 
was discussed in  \ci{TH}).}
\be
 \A_\vp =  {\g \ov r^2} \sin^2 \t \ , \ \ \ \ \ \ 
\A_\psi =  {\g \ov r^2} \cos^2 \t , \ \ \ \ \ \ \g=\const .  
\la{sool}
\ee
[ In the $SO(3)$-invariant case \re{larr}
(related to the $D=4$ dyonic solution)  one finds  that  the self-duality 
condition becomes (we assume that the fields do not depend on the internal coordinate $y_1$)\foot{Here
$H_{y_1pq}= \del_q b_p - \del_p b_q, \ G^{y_1y_1} = f\inv (k\inv + k a_s a_s), \ 
G^{pq}= f\inv k \delta^{pq}, \ G^{py_1}= - f\inv k a_p$, 
\ $\sqrt G= f^2 k\inv$, \ $p,q,..=1,2,3$. All repeated indices are contracted with flat metric. }
 $$  \F_{pq} + a_p \del_q A - a_q \del_p A  +  k\inv \ep_{pqs} \del_s A  =0 \ , \ \ \ \ A(x) \equiv \A_{y_1} \ . $$
Since $\del_{p} a_q - \del_q a_{p}
 = - \epsilon_{pqs} \del_s   k\inv $ this becomes
$ \del_p  \td \A_q$ $ - \del_q  \td \A_p= - k^{-2} \ep_{pqs}
  \del_s (k A) ,
$ where $ \td \A_p $ $ \equiv  \A_p - Aa_q   . $
Then $\del^s[k^{-2} \del_s (kA)]=0$, i.e.  $A(r) = q/r$ 
and $ \td \A_\vp = q (1-\cos \theta)$ in the simplest 
1-center case, so that  we  get  an extra Taub-NUT  term in the resulting 
 $D=4$  metric (i.e. \re{eee} with $dt^2 \to (dt + \A_\vp d\vp)^2$
and $\l^{-2} \to F\inv K f k\inv - k^{-2}A^2$, cf.\re{mag}).
It is  interesting  that all the charges can be superposed
 in a way consistent with conformal invariance 
with the  Taub-NUT charge  being related to the charge of $A$. 
The solution \re{gene}  found in \ci{USS}
corresponds to  $\td \A_p =0$ but with 
$A$  subject not to \re{ins} but  the full second order equation \re{sell}.
The $n=y_1$ component of it reduces to
 $\del^s[f\inv k^{-3}\del_s (k A)]=0$  (which is solved by \re{gene}) 
while  three other components  are satisfied identically. ]

Since the presence of $\A_m$-coupling does not change the equations
\re{LL}  for 
the functions $F,K$ they can  still  be chosen as in  \re{ooo}.
Dimensionally reducing this  $D=6$ model \re{laggg},\re{hppp} to 5 dimensions along $u$ as in Section 2.3 
one finds  the 
rotating  generalisation of the 3-charge $(Q_1,Q_2,P)$  extreme dyonic $D=5$ black hole background \re{eeee}. 
The resulting  5-dimensional Einstein-frame metric 
is 
\be
 ds^2_E =- \l^{2} (r)  (dt + \A_m dx^m)^2 +  \l\inv  (r)  dx^m dx_m 
\la{eye}
\ee
$$
=  - \l^{2} (r) 
 [dt  + {\g \ov r^2}( \sin^2 \t  d\vp  +  \cos^2 \t d\psi)]^2 
+  \l\inv  (r) [ dr^2 + r^2 (d\theta^2 +
 \sin^2\theta d\vp^2  + \cos^2\t d\psi^2)]   , 
$$
$$
\l= {r^{2} \ov [(r^2+Q_1)(r^2+Q_2)(r^2+ P)]^{1/3}}\ , 
$$ 
where $\l(r)$ is the same as in  the static case \re{eeee}.
As for $\g=0$ the surface  $r=0 $ is a regular horizon.
The mass of  this  $D=5$ rotating dyonic black hole 
is  still given by \re{maa}  while the angular momenta in the two planes are\foot{In general, for a black hole in  $D$ dimensions with the metric 
$\ g_{ti} =   {\g\ep_{ik}  x^k /r^{D-2}}  + ...$ 
the  angular momentum in $(i,k)$ plane is  \ci{MPE}  
 $ J= \g \om_{D-2}/8 \pi G^{(D)}_N$.}
\be
 J_\vp =  J_\psi= J= {\pi \ov 4G_N} \g  \ . 
\la{momm}
\ee
In the  $r\to 0$ limit  the metric  becomes 
$$
 (ds^2)_{r\to 0} = (Q_1Q_2P)^{1/3}  [ (d \ln r)^2 + d\t^2 +  \sin^2 \t 
  d\vp^2  +   \cos^2\t d \psi^2  $$ 
\be 
  -  \ { \g^2\ov Q_1Q_2P} 
(\sin^2 \t   d\vp  +   \cos^2\t d \psi)^2
] \  ,  
\la{mmmn}
\ee
leading  to the following expression for the area of the $r=0$ horizon 
(cf.\re{amaa})
\be
  {\bf A} = 2 \pi^2 \sqrt{ Q_1Q_2P - \g^2} \ .   
\la{aar}
\ee
The combination $ Q_1Q_2P - \g^2$ should be positive in order 
for the signature of the metric \re{mmmn} to remain euclidean.
Thus there is no regular rotating solution in the 
electric limit $P\to 0$.

Expressed in terms of quantised charges \re{chaa}  and the  
angular momentum \re{momm} the 
 entropy  becomes
\be 
{\bf S}_{BH}=  {{\bf A}\ov 4G_N}=  {\pi^2 \sqrt{ P Q_1 Q_2  - \g^2 } \ov  2 G_N} 
  = 2  \pi \sqrt{ \k mw  -  J^2} \ . 
\la{etep}
\ee
The special case $Q_1=Q_2=P$ of this solution  
(written in terms of  $\rho$ in \re{eeew}) 
 was found in \ci{MV} where the  corresponding 
thermodynamic entropy was also obtained using $D$-brane counting 
approach. Interpreting the above general  $D=6$  background
as a solution of type IIB theory in 10 dimensions
and applying $SL(2,Z)$-duality to transform it 
into a solution supported by RR-charges it should be straightforward 
to reproduce  the entropy  \re{etep} by generalising 
the counting done in the non-rotating case in \ci{CM}. 

%%%%%%%%%%%%%%%%%%%%%%%%%%%%%%%%%%%%%%%%%%%%%%%%%%%%%%%%%
\subsection{Throat limit,  bound on angular momentum and entropy }
%%%%%%%%%%%%%%%%%%%%%%%%%%%%%%%%%%%%%%%%%%%%%%%%%%%%%%%%%%%
Having  identified  a conformal \sm 
behind this rotating solution one is able to  derive some of  its properties  directly from    the corresponding conformal theory.
As in  the non-rotating case \re{nnn}  the 
throat limit $r\to 0$  of this  model 
is {\it regular} and is described
by   $SL(2,R) \times SU(2)$ 
WZW  theory.
The angles $(\t$,$\vp$,$\psi)$ in \re{ccr}  
are  related to the standard $S^3$ Euler angles used in \re{nnn} 
 by (these Euler  angles in \re{nnn}   here will be denoted by primes)
$$\t= \ha \t', \ \ \ \vp= \ha (\vp' + \psi'), \   \ \ \psi=  \ha (\psi' - \vp'), \ \ \  0 \leq \t' \leq \pi, \   0 \leq \vp' \leq 2\pi,\ 
0 \leq \psi' \leq 4\pi.$$
Defining  $z \equiv \ln {Q_2 \ov  r^2}$ we get  (cf.\re{nnn})
$$
 I_{r \to 0}  
=  {1 \ov \pi\a' }\int d^2 \s \ 
 \left[ e^{-z }   \del   u \bd  v  + Q_1  Q_2\inv \del   u  \bd   u 
+   \g Q_2\inv \del u  (\bd \psi' -    \cos\t' \bd \vp')\right]  $$
\be
  + \ 
{P \ov 4\pi \a' }\int d^2 \s  \left(\del z \bd z
 +  \del \psi' \bd   \psi'  +  \del \vp' \bd \vp' + 
\del \theta' \bd \theta'  - 2 \cos \theta'  \del  \psi' \bd \vp'    \right)  .  \la{nnne}
\ee
The $\g$-term  ($\sim  \del u \bar J_3$)
 can be interpreted as an integrable marginal deformation
of the $SL(2,R) \times SU(2)$ WZW model (similar models were discussed in \ci{kirr,Tt}).
Introducing 
$$\psi'' = \psi' + 2\g  P\inv Q_2\inv u$$
we can represent this action in the   $SL(2,R) \times SU(2)$ WZW form as in \re{nnn} 
\be
 I_{r \to 0}  
=  {1 \ov \pi\a' }\int d^2 \s \left[ e^{-z }   \del   u \bd  v 
 +  (PQ^2_2)\inv   (Q_1 Q_2 P - \g ^2) \del   u  \bd   u  \right] 
\la{nnen}
\ee
$$
  + \ 
{P \ov 4\pi \a' }\int d^2 \s  \left(\del z \bd z
 +  \del \psi'' \bd   \psi''  +  \del \vp'\bd \vp' + 
\del \theta' \bd \theta'  - 2 \cos \theta'  \del  \psi'' \bd \vp'    \right)  .  $$
The two  WZW models (described by $(z,u,v)$ and $(\theta',\vp',\psi'')$) have the same level 
 $\k = P/\a'$.
 The consequences of this representation are:

(1) Quantization of parameters: \ 
 $P=\a' \k $ is quantized since the level of $SU(2)$ is. 
 Demanding  that $\psi''$  should have  the same  ($4\pi$)
periodicity 
as
$\psi'$   we conclude that  the value of $\g$ should be quantised
$  \g  R P\inv Q_2\inv  =  l=$integer ($R$ is the radius of the compact coordinate $u$).\foot{From the low-dimensional classical effective action point of view $\g$ can, of course,  take   continuous set of values.
In principle, it is not necessary to  insist on representing \re{nnne}
in the `factorised' $SL(2,R)\times SU(2)$ form \re{nnen}
with $4\pi$-periodic $\psi''$. However, if $\g$ is not quantised 
it  breaks space-time supersymmetry at the string-theory level
(as in the `magnetic' models in \ci{rus,Tt,kirr}).
Note that like 
(world-sheet supersymmetric)  
$SL(2,R)\times SU(2)$ WZW model the theory \re{nnne} has 
zero  central charge deficit. Continuous supersymmetry breaking does not 
contradict standard lore since this model is non-compact.
Supersymmetry is important  in the present context in order to  be able to  reproduce  
the entropy \re{etep} as a statistical one (it is only if $J$ is quantised that 
$\k mw -J^2$ may be  related to  a number of microscopic  states).}
Relating $Q_2$ to the  `winding number' $w$   \re{chaa} 
 and  $\g$ to $J$ \re{momm}  we conclude  that 
 \be
 P= \a' \k, \ \ \  \ \g= P {Q_2\ov R} l , \ \  \  i.e.  \  \  \
J= \k w l , 
\la{bgb}
\ee
so that $J$ should  take only integer values.

(2) Bound on  angular momentum: \ 
In order for $u$  to have positive norm  in \re{nnen}
we should have  (we assume  that all the charges are positive) 
 \be
 Q_1 Q_2 P - \g ^2  = ({4G_N \ov \pi})^2 (\k mw - J^2) > 0, \ \  \ 
i.e.  \ \ \ 
  J^2  < \k mw \ , 
\la{bou}
\ee 
getting a constraint on maximal  
 value of $\g $ or $J$.  
This `Regge bound'   thus follows directly  from the 
regularity of   
underlying conformal  $\s$-model.\foot{The same condition is necessary to avoid   closed time-like geodesics in the resulting  dimensionally reduced ($D=5$) metric \ci{MV}.}

The  structure of the throat region theory  and the 
fact that  the angular momentum terms ($\A_i \del u \bd x^i$) 
 appear in the conformal \sm 
action \re{laggg} in the same  way as (and indeed are particular examples  of)
 marginal perturbations 
corresponding to left-moving oscillations 
of the solitonic string (`fundamental string wound around 5-brane')
strongly suggests a possibility to understand the expression for the entropy 
\re{etep}  along the same lines as in the non-rotating case discussed
in Section 3.  Introducing a $u$-dependence into $\g$
or $J$, \ $J(u)= J + \td J(u)$,  and treating  $J(u)$ as a perturbation 
of the  original model\foot{The perturbation 
$\sim \g(u)\del u  (\bd \psi' -    \cos\t' \bd \vp')$ of $SL(2,R) \times SU(2)$ WZW theory  (cf.\re{nnne}) remains  marginal for arbitrary
function $\g(u)$ (as follows from 
 the presence of the mixed $\del u \bd v$-term 
in $SL(2,R)$ action).}
one should get, using $1/\k$ expansion, 
 the  following  level matching condition
\be 
mw =  {1\ov \k} < J^2(u)> ={1\ov \k} J^2 
+ {1\ov \k} < \td J^2(u)>, \ \ \ \ 
\ N_L \sim < \td J^2(u)> = \k mw -J^2  .  
\la{ttti}
\ee
This argument is exactly parallel to the one given 
in Section 3 with $J(u)$ playing the role of a `charge'  $q(u)$ (cf.\re{aap}--\re{ijn}).
As in \re{eeen} the corresponding statistical entropy then reproduces 
the expression for the Bekenstein-Hawking one \re{etep}.

%%%%%%%%%%%%%%%%%%%%%%%%%%%%%%%%%%%%%%%%%%%%%%%%%%%%%%%%%
%\subsection{More general solutions
%with two independent components of rotation
%}
%%%%%%%%%%%%%%%%%%%%%%%%%%%%%%%%%%%%%%%%%%%%%%%%%%%%%%%%%%%
Let us finish with remarks 
on other related solutions.
 The $D$-brane counting arguments
in \ci{MV}  imply  that there  should exist a solution
with independent values of angular  momenta  $J_\vp$ and $J_\psi$ 
in the two planes. Such a solution  is
 not described by the 
chiral null model \re{laggg} (which, at the same time,  represents almost all  known  exact extreme solutions).\footnote{In particular,  a solution with an angular momentum in only one of the 
two planes  (e.g. $\A_\vp\not=0, \ \A_\psi=0$)
must satisfy, according to \re{sell},\re{FFF} 
 $\del_r f\inv \del_\t \A_\vp=0$ so that 
$\A_\vp=\A_\vp(r)$ if $P\not=0$.
Then $ \A_\vp = - c \ln r + {cP \ov 2r^2}, \ a=\const,$
which  does not decay at large $r$.}
% This may not be  not surprising since $|J_\vp| \not= |J_\psi|$
%background is   expected to  have one half less of supersymmetry \ci{MV}.
One may also consider a generalisation 
to the case when the harmonic functions $F\inv,K,f$
are not spherically but only axially symmetric
($x^1+ix^2 = \sqrt{r^2 + a^2} \sin \theta e^{i \vp}$, 
\  $F\inv = 1 +  {Q_2  \over r^2  + a^2 cos^2 \theta}$, etc.).
%\A_\vp = -{\g   sin^2\theta  \over r^2  + a^2 cos^2 \theta}$,  
It may be of interest  to construct rotating $D=4$ solutions
by reducing these   $D=5$ solutions (their multicenter generalisations) 
down to 4 dimensions  as in 
\ci{HS,HH}.

%%%%%%%%%%%%%%%%%%%%%%%%%%%%%%%%%%%%%%%%%%%%%%%%%%%%%%%%%%%%%%
 \section*{Acknowledgments}
I am  grateful to M. Cveti\v c  for collaboration on \ci{US,USS}  and useful discussions. 
I would like  to thank  also G. Gibbons, F. Larsen and T. Ort\'{\i}n 
for  interesting   questions and remarks.
This work was  supported by  PPARC,   
 ECC grant SC1*-CT92-0789 and NATO  grant CGR 940870.

%%%%%%%%%%%%%%%%%%%%%%%%%%%%%%%%%%%%%%%%%%%%%%%%%%%%%%%%%%%%

% \section*{References}

%%%%%%%%%%%%%%%%%%%%%%%%%%%%%%%%%%%%%%%%%%%%%%%%%%%%%%%%%%%%%
%%%%%%%%%%%%%%%%%%%%%%%%%%%%%%%%%%%%%%%%%%%%%%%%%%%%%%%%%%%%%%
 

\begin{thebibliography}{99}
\bibitem{kall}{R. Kallosh, A. Linde, T. Ort\' in, A. Peet and A. van Proeyen, \pr {\bf D}46 (1992) 5278.} 

\bibitem{HRT}{G.T. Horowitz and A.A. Tseytlin,
 \prl {\bf 73} (1994) 3351, hep-th/940840.}

\bibitem{duh}{A. Dabholkar and J.A. Harvey, \prl {\bf 63} (1989) 478;
 A. Dabholkar, G.W.   Gibbons, J.A.   Harvey  and F. Ruiz-Ruiz,
\np {\bf B340} (1990) 33. }

\bibitem{CMP}{ C.G. Callan, J.M.  Maldacena  and A.W. Peet, 
%``Extremal black holes as fundamental strings", 
PUPT-1565,  hep-th/9510134.  } 

\bibitem{dab}{  A. Dabholkar, J.P. Gauntlett, J.A. Harvey and D. Waldram, 
%``Strings as solitons and black holes as strings",
 CALT-68-2028, hep-th/9511053.   }

\bi{DRU}{M.J. Duff and  J. Rahmfeld, \pl 
{\bf B345} (1995) 441,  hep-th/ 9406105.} 

\bibitem{seen} {A. Sen, \np {\bf B440} (1995) 421, hep-th/9504027. }

\bi{suss} {L. Susskind, RU-93-44, hep-th/9309145;  J. Russo and L. Susskind, \np {\bf B}437 (1995) 611,  hep-th/9405117.}

\bibitem{ME}{ A.A. Tseytlin,  \pl  {\bf B}363 (1995) 223,  hep-th/9509050. }

\bibitem{sen}{A. Sen, \mpl  {\bf A10} (1995) 2081, 
 hep-th/9504147. }

\bibitem{HMON}{R. Khuri, \pl  {\bf B259} (1991) 261; \np {\bf B387} (1992) 315; J. Gauntlett, J.A. Harvey and J. Liu, 
\pl {\bf B409} (1993) 363.}

\bibitem{kalor}{W. Nelson, \pr {\bf D49} (1994) 5302; R. Kallosh  and  T. Ort\'{\i}n, \pr  {\bf D50} (1994)
7123, hep-th/9409060.}


\bibitem{peet}{A.W.  Peet, \np  {\bf B456} (1995) 732, 
 hep-th/9506200.}


\bibitem{CY}{M. Cveti\v c and D. Youm,
 UPR-0672-T, hep-th/9507090; UPR-0675-T, hep-th/9508058; 
  \pl {\bf B359} (1995) 87, 
hep-th/9507160.}

\bi{KOO}{T. Ort\'{\i}n,  \pr {\bf D47} (1993) 3136; 
 hep-th/9208078; R. Kallosh and T. Ort\'{\i}n,
 \pr {\bf D48} (1993) 742; 
 hep-th/9302109.}
\bibitem{IWP}{R. Kallosh, D. Kastor, T. Ort\'{\i}n and T. Torma, 
\pr {\bf D50} (1994) 6374, hep-th/9406059.}

\bi{KHO}{R.R. Khuri and T. Ort\'{\i}n, CERN-TH/95-347, hep-th/9512178.}

\bibitem{US}{M. Cveti\v c and  A.A.  Tseytlin, 
\pl {\bf B366} (1996) 95, hep-th/9510097. 
}
\bibitem{USS}{M. Cveti\v c and  A.A.  Tseytlin, 
IASSNS-HEP-95-102, hep-th/9512031. 
}
\bibitem{LW}{ F. Larsen  and F. Wilczek, 
%``Internal structure of black holes", 
PUPT-1576,  hep-th/9511064.    }


\bi{SV} {A. Strominger and C. Vafa, HUTP-96-A002,  hep-th/9601029.}

\bi{MV} {J.C. Breckenridge, R.C. Myers, A.W. Peet  and C. Vafa, HUTP-96-A005,  hep-th/9602065.}


\bibitem{TS}{A.A. Tseytlin,  \cqg {\bf 12}  (1995) 2365, hep-th/9505052.}

\bibitem{TH}{ G.T. Horowitz and A.A. Tseytlin,  \pr {\bf D51} (1995) 
2896, hep-th/9409021;   \pr {\bf D50} (1995) 5204, hep-th/9406067.}

\bi{GIBMA}{G.W.  Gibbons,   \np {\bf B207}  (1982) 337;
G.W.  Gibbons and K. Maeda, \np {\bf  B298} (1988) 741; D. Garfinkle, G.T.  Horowitz and A. Strominger,  \pr {\bf D43} (1991)
3140; {\bf D45} (1992) 3888 (E).}

\bibitem{kalorr}
{K. Behrndt and R. Kallosh, SU-ITP-95-33, hep-th/9601010;
 SU-ITP-95-19, hep-th/9509102;
}

\bi{chs}{C.G. Callan, J.A. Harvey and A. Strominger, 
\np {\bf B359 } (1991)  611; in {\it 
Proceedings of the 1991 Trieste Spring School on String Theory and
Quantum Gravity}, J.A. Harvey {\it et al.,}  eds. (World Scientific, 
Singapore
1992).}

\bi{behh}{ K. Behrndt, \np  {\bf B455} (1995) 188, hep-th/950610.}
\bi{behhh}{
K. Behrndt and H. Dorn, hep-th/9510178; K. Behrndt, E. Bergshoeff and B. Janssen, 
UG-13/95, hep-th/9512152.}

\bibitem{RRR}{J. Rahmfeld, CTP-TAMU-51/95,   hep-th/9512089;
 T. Ort\'{\i}n,  CERN-TH/96-36, hep-th/9602067. }

\bibitem{DLR}{M.J. Duff, J.T. Liu and J. Rahmfeld, CTP-TAMU-27-95,
 hep-th/9508094.} 



\bi{ENT}{S. W. Hawking, G.T. Horowitz and S.F. Ross, 
\pr {\bf D51} (1995) 4302;
C. Teitelboim, \pr {\bf D51} (1995) 4315; 
G.W. Gibbons and R.E. Kallosh, \pr{\bf  D51} (1995) 2839.}

\bi{gho}{A. Ghosh and P. Mitra, hep-th/9602057.}
\bibitem{HP}{P. Howe and G. Papadopoulos, \np  {\bf B289} (1987) 264;   {\bf B381} (1992) 360. }

\bibitem{DL}{M.J.  Duff and J.X. Lu,  \np {\bf B354}  (1991) 141.}


\bibitem{MK}{R. Myers and R. Khuri, hep-th/9512061. }


\bibitem{dukh} {A. Sen,  \np {\bf B450} (1995) 103,
hep-th/9504027; J.A.  Harvey  and A. Strominger,  \np {\bf B449 }
(1995) 535, hep-th/9504047.}

\bibitem{duf}{M.J.  Duff, S. Ferrara, R.R. Khuri and J. Rahmfeld, \pl {\bf  B356} 
(1995) 479, hep-th/9506057.}

\bi{MPE}{R. Myers and M. Perry, \ap {\bf 172 }  (1986) 304.}
\bibitem{CYY}{M. Cveti\v c and D. Youm, IASSNS-HEP-95-107, hep-th/9512127; 
 D. Jatkar, S. Mukherji and S. Panda, hep-th/9512157. }

\bi{SL}  {D.A. Lowe and A. Strominger, \prl {\bf 73} (1994) 1468, hep-th/9403186.}

\bibitem{CM}{ C.G. Callan and  J.M.  Maldacena, 
PUPT-1591,  hep-th/9602043.} 

\bi{kirr}{E. Kiritsis and C. Kounnas, \np {\bf B456} (1995) 699, hep-th/9508078.}
\bi{Tt}{A.A. Tseytlin, Imperial/TP/94-95/62, hep-th/9510041.}

\bi{rus}{J.G. Russo and A.A. Tseytlin, CERN-TH/95-215, hep-th/9508068.}

\bi{HS}{G.T.  Horowitz and  A. Sen, \pr {\bf D53} (1996) 808, 
 hep-th/9509108.}
\bi{HH}{G.T.  Horowitz and  T. Tada,  UCSBTH-95-39, hep-th/9601004.}


\end{thebibliography}
\end{document}